\documentclass[english,sort,aps,prd,reprint,nobibnotes,notitlepage,nofootinbib,showpacs,preprintnumbers]{revtex4-1}
\usepackage[latin9]{inputenc}
\usepackage{geometry}
\usepackage{fancyhdr}
\usepackage{color}
\usepackage{babel}
\usepackage{amsmath}
\usepackage{amssymb}
\usepackage{graphicx}
\usepackage{esint}
\usepackage[unicode=true,
 bookmarks=true,bookmarksnumbered=false,bookmarksopen=false,
 breaklinks=false,pdfborder={0 0 1},backref=false,colorlinks=true]
 {hyperref}
\hypersetup{pdftitle={How Many e-Folds Should We Expect from High-Scale Inflation?},
 pdfauthor={Grant N. Remmen & Sean M. Carroll},
 citecolor=blue,linkcolor=blue,urlcolor=blue,citecolor=blue,linkcolor=blue}
\usepackage{breakurl}

\makeatletter
 
 \@ifundefined{textcolor}{}
 {%
   \definecolor{BLACK}{gray}{0}
   \definecolor{WHITE}{gray}{1}
   \definecolor{RED}{rgb}{1,0,0}
   \definecolor{GREEN}{rgb}{0,1,0}
   \definecolor{BLUE}{rgb}{0,0,1}
   \definecolor{CYAN}{cmyk}{1,0,0,0}
   \definecolor{MAGENTA}{cmyk}{0,1,0,0}
   \definecolor{YELLOW}{cmyk}{0,0,1,0}
 }

\usepackage{babel}
\@ifundefined{definecolor}{color}{}
\usepackage{babel}

\usepackage{fancyhdr}\@ifundefined{definecolor}{color}{}
\usepackage{babel}

\usepackage{xcolor}
\usepackage{fancyheadings}
\makeatother

\begin{document}

\preprint{\hbox{CALT-TH-2014-138} }

\title{How Many $e$-Folds Should We Expect from High-Scale Inflation?}

\author{Grant N. Remmen}

\email{gremmen@theory.caltech.edu}

\author{Sean M. Carroll}

\email{seancarroll@gmail.com}

\date{\today}

\affiliation{Walter Burke Institute for Theoretical Physics,\\
 California Institute of Technology, Pasadena, CA 91125}

\begin{abstract}
We address the issue of how many $e$-folds we would naturally expect
if inflation occurred at an energy scale of order $10^{16}$\,GeV.
We use the canonical measure on trajectories in classical phase space,
specialized to the case of flat universes with a single scalar field.
While there is no exact analytic expression for the measure, we are
able to derive conditions that determine its behavior. For a quadratic
potential $V(\phi)=m^{2}\phi^{2}/2$ with $m=2\times10^{13}$\,GeV
and cutoff at $M_{{\rm Pl}}=2.4\times10^{18}$\,GeV, we find an expectation
value of $2\times10^{10}$ $e$-folds on the set of Friedmann--Robertson--Walker trajectories.
For cosine inflation $V(\phi)=\Lambda^{4}[1-\cos(\phi/f)]$ with $f=1.5\times10^{19}$\,GeV,
we find that the expected total number of $e$-folds is $50$, which
would just satisfy the observed requirements of our own Universe;
if $f$ is larger, more than $50$ $e$-folds are generically attained.
We conclude that one should expect a large amount of inflation in
large-field models and more limited inflation in small-field (hilltop)
scenarios. 
\end{abstract}
\maketitle

\section{Introduction}

The possible detection of tensor perturbations in the cosmic microwave
background (CMB) by the BICEP2 experiment \cite{Bicep2} suggests
that inflation occurred at a high energy scale \cite{Baumann}: $E_{\mathrm{I}}=2\times10^{16}$~GeV,
just two orders of magnitude below the reduced Planck scale $M_{\mathrm{Pl}}=1/\sqrt{8\pi G}=2.4\times10^{18}$~GeV.
Knowing this parameter with some confidence allows both for much more
focused inflationary model-building and for quantitative exploration
of some of the conceptual issues underlying the inflationary paradigm.
In this paper, we address one of the latter: given an inflaton potential
that is able to reproduce the measured cosmological parameters, how
much inflation is likely to have occurred? In the present work, we answer
this question, finding that the expected number of $e$-folds of inflation
depends dramatically on the general type of inflaton potential chosen.

The amount of inflation that occurs is measured by the number of $e$-folds,
\begin{equation}
N=\int_{a_{{\rm i}}}^{a_{{\rm f}}}\mathrm{d}\ln{a}=\int_{t_{{\rm i}}}^{t_{{\rm f}}}H\,\mathrm{d}t.\label{efoldsdefined}
\end{equation}
Here, $a_{{\rm i}}$ and $a_{{\rm f}}$ are the values of the scale
factor at the beginning and end of inflation, while $t_{{\rm i}}$
and $t_{{\rm f}}$ are the corresponding proper times. We can define
the period during which inflation is occurring as that for which the
Universe is accelerating, $\ddot{a}>0$. In conventional inflationary
models, it is necessary to achieve at least 50 $e$-folds to successfully
address the horizon problem. It is generally accepted that this requirement
can be met by a wide variety of potentials.

We would like to know not only whether a certain potential can possibly
produce sufficient amounts of inflation, but whether such an outcome
is actually likely. Presumably, a complete theory of cosmological
initial conditions in the context of quantum gravity would provide
a unique answer to this question, but we don't have such a theory
at present. What we do have are classical models of inflaton dynamics
coupled to general relativity. Any classical theory comes with a natural
measure on phase space, the Liouville measure. Gibbons, Hawking, and
Stewart (GHS) showed how to use this measure to define a canonical
measure on cosmological \emph{trajectories} (rather than individual
points in phase space) \cite{GHS}. In this measure, we can calculate
the fraction of universes with given properties, such as ``more than
50 $e$-folds of inflation.'' Given the current state of the art,
this is the best we can do to decide whether such solutions are likely
or not.

The GHS measure comes with a technical problem when applied to (homogeneous,
isotropic) Friedmann--Robertson--Walker (FRW) cosmologies: it diverges
as the spatial curvature approaches zero, assigning almost all measure
to flat universes. Different proposals have been advanced for dealing
with this divergence, including removing the region of infinite measure
by hand \cite{GibbonsTurok}. As noted in Refs.~\cite{CarrollTam}
and \cite{attractors}, the divergence for flat universes is an indication
that, in the canonical measure, almost all cosmological spacetimes
are flat. For this reason, and also given the physical relevance of
spatially flat solutions \cite{Planck}, it is on these that we concentrate
our efforts. In a previous paper \cite{attractors}, we developed
a formalism for defining the Hamiltonian-conserved measure on the
effective two-dimensional phase space for a canonical scalar field
with a potential in a flat FRW cosmology. Although we did not prove
the uniqueness of this measure in arbitrary theories, we could establish
it for quadratic potentials and expect it to hold for well-behaved
potentials more generally.

In this paper, we employ the formalism developed in Ref.~\cite{attractors}
to study high-scale inflation. We focus on two representative models:
quadratic inflation and cosine (``natural'') inflation. We find
dramatically different quantitative results for the two cases. In
quadratic inflation, given that the potential is chosen to fit observed
cosmological parameters, we find that large amounts of inflation are
favored by the canonical measure --- billions of $e$-folds of inflation
--- provided we extrapolate the quadratic potential up to the Planck
scale $H=M_{\mathrm{Pl}}$ and allow the inflaton field $\phi$ to
run over a super-Planckian range $\sim10^{5}M_{\mathrm{Pl}}$. Moreover,
we find that almost all trajectories experience well more than 50
$e$-folds. For cosine potentials, by contrast, the expected amount
of inflation under the canonical measure is relatively small: if the
symmetry-breaking parameter $f$ is set to the reduced Planck scale,
$M_{\mathrm{Pl}}=2.4\times10^{18}$~GeV, we expect of order one $e$-fold,
with the probability of attaining as many as 50 $e$-folds being exponentially
small. These numbers depend sensitively on $f$; once it is above
$10^{19}$~GeV, as favored by the BICEP2 result \cite{Freese,Bicep2},
the probability of getting more than 50 $e$-folds rises above 50\%.

This last result is interesting, since cosine potentials feature ``hilltops''
from which trajectories with arbitrarily large numbers of $e$-folds
can originate. Our analysis demonstrates that, while such lingering
solutions are allowed, they contribute a relatively small amount to
the measure on the space of trajectories. We conjecture that this
behavior reflects a more general difference between potentials that
rise up to the Planck scale, in which we expect large amounts of inflation,
and models with potential maxima below the Planck scale, where the
expected number of $e$-folds will be comparatively small.

Any analysis of this form necessarily comes with caveats. As noted,
we are using a classical measure, whereas a particular theory of initial
conditions (\textit{e.g.}, a proposal for the wave function of the
Universe) will presumably make its own predictions. More seriously,
our analysis applies only to universes that are assumed to be homogeneous
from the start. Once perturbations are included, it is clear that
most universes should be wildly inhomogeneous; the existence of the
sufficiently smooth initial conditions necessary for inflation to
begin is highly non-generic \cite{CarrollTam,Penrose}. Given the
evidence that inflation did happen, we consider the expected number
of $e$-folds according to the canonical measure to be a useful diagnostic
of which models are robust and which are more delicate. An ultimate
justification for why inflation occurs in the first place awaits further
insight.

This paper is organized as follows. In Sec.~\ref{Review} we first
review the formalism of Refs.~\cite{GHS} and \cite{attractors}
for finding the canonical measure on phase space, as well as the sense
in which phase space becomes effectively only two-dimensional for
flat FRW cosmologies. The connection between the measure on effective
phase space and the measure on the space of possible trajectories
of evolution of a FRW universe is presented. Next, in Sec.~\ref{Generic}
we derive some general properties of the measure for arbitrary slow-roll
and hilltop potentials. Finally, we examine representative models
of each class, quadratic inflation and cosine inflation, in Secs.
\ref{Quadratic} and \ref{Cosine}, making statistical calculations
on the ensemble of all FRW universes and finding the expected number
of $e$-folds of inflation attained.

\section{The Probability Distribution on the Set of Universes}

\label{Review}

\subsection{The Hamiltonian-conserved measure}

We are interested in the theory of a homogeneous scalar field in an
expanding FRW universe. The action is 
\begin{equation}
S=\int {\rm d}^{4}x\,\sqrt{-g}\left[\frac{M_{\mathrm{Pl}}^{2}}{2}R-\frac{1}{2}g^{\mu\nu}\partial_{\mu}\phi\partial_{\nu}\phi-V(\phi)\right].
\end{equation}
The metric can be written 
\begin{equation}
\mathrm{d}s^{2}=-N^{2}(t)\mathrm{d}t^{2}+a^{2}(t)\left(\frac{\mathrm{d}r^{2}}{1-\kappa r^{2}}+r^{2}\mathrm{d}\Omega^{2}\right),
\end{equation}
where $N$ is the lapse function and the curvature parameter $\kappa$
is an arbitrary real parameter with mass dimension $2$. The number
$\kappa$ is fixed for a given FRW universe and we can write $k=\kappa R_{0}^{2}\in\left\{ -1,0,1\right\} $,
where $R_{0}$ is the radius of curvature of the universe at unit
scale factor. Taking $\phi(t)$ to depend only on time, the Hamiltonian
is 
\begin{equation}
\begin{aligned}\mathcal{H} & =-3a^{3}NM_{\mathrm{Pl}}^{2}\!\left\{ \!\frac{\dot{a}^{2}}{a^{2}}\!+\!\frac{\kappa}{a^{2}}\!-\!\frac{1}{3M_{\mathrm{Pl}}^{2}}\left[\!\frac{1}{2}\dot{\phi}^{2}\!+\! V(\phi)\!\right]\!\right\} \\
 & =N\left[-\frac{p_{a}^{2}}{12aM_{\mathrm{Pl}}^{2}}+\frac{p_{\phi}^{2}}{2a^{3}}+a^{3}V\left(\phi\right)-3a\kappa M_{\mathrm{Pl}}^{2}\right],
\end{aligned}
\end{equation}
where $p_{a}$ and $p_{\phi}$ are the momenta conjugate to the scale
factor and scalar field, respectively. The scalar equation of motion
is 
\begin{equation}
\ddot{\phi}+3H\dot{\phi}+V'(\phi)=0,\label{KG}
\end{equation}
where $V^{\prime}(\phi)=\mathrm{d}V/\mathrm{d}\phi$. The Hamiltonian
constraint, which comes from varying with respect to $N$, is the
Friedmann equation, 
\begin{equation}
H^{2}=\frac{1}{3M_{\mathrm{Pl}}^{2}}\left[\frac{1}{2}\dot{\phi}^{2}+V\left(\phi\right)\right]-\frac{\kappa}{a^{2}},\label{Hubble}
\end{equation}
where $H=\dot{a}/a$ is the Hubble parameter.

Any classical theory comes with a preferred choice of measure on phase
space: the Liouville measure, which is preserved under time evolution.
In cosmology our interest is less in a measure on individual points
in phase space and more in a measure on trajectories through time
or specific cosmological evolutions. Gibbons, Hawking, and Stewart
\cite{GHS} showed how to construct such a measure for a scalar field
coupled to general relativity. The phase space is na\"{i}vely four-dimensional,
with coordinates given by $a$ and $\phi$ and their conjugate momenta.
But the Hamiltonian constraint, implemented by the Friedmann equation,
cuts this down to three dimensions. The space of trajectories (equivalent
under the equations of motion to the space of initial conditions)
is one lower, leaving us with a two-dimensional space. GHS were able
to construct a unique measure on this space that is positive and invariant
under time evolution (for further discussion see Refs.~\cite{GibbonsTurok,CarrollTam,attractors}).

As Ref.~\cite{CarrollTam} shows, the GHS measure \cite{GHS} has
an interesting property: on a transverse surface in phase space defined
by fixed Hubble parameter, the measure diverges for small curvature
$\kappa$ as $\left|\Omega_{k}\right|^{-5/2}$. This behavior has
the good feature that it implies that the collection of non-flat FRW
universes is a set of measure zero under the GHS measure; that is,
the flatness problem in cosmology is solved by the GHS measure, since
almost all trajectories are flat. However, from the point of view
of understanding the set of flat FRW universes itself, this behavior
poses a technical challenge. It is difficult to regularize the divergence
in the GHS measure to construct a well-defined measure within the
space of flat universes.

In our previous paper, we showed how to find a measure on the space
of flat universes by constructing it by hand, subject to the requirement
that it be conserved under time evolution \cite{attractors}. We note
from Eqs.~\eqref{KG} and \eqref{Hubble} that the scale factor $a$
disappears from the equations of motion when $\kappa=0$. The effective
phase space is therefore only two-dimensional; specifying the two
quantities $\phi$ and $\dot{\phi}$ completely determines the solution
(although they are not conjugate variables). The set of trajectories
in effective phase space is therefore one-dimensional. In Ref.~\cite{attractors}
we formalized the notion of an effective phase space via the property
of vector field invariance between two manifolds. We argued that there
exists a unique measure on this space that is conserved under Hamiltonian
flow, in analogy with the conventional Liouville measure, which one can use to construct a measure on the space of flat universes.

The time evolution given by Eq.~\eqref{KG} can be characterized by
a vector field $\mathbf{v}$ on $\phi$-$\dot{\phi}$ space, with
components 
\begin{equation}
\mathbf{v}=\left(\dot{\phi},-V^{\prime}\left(\phi\right)-3H\dot{\phi}\right).\label{velocity}
\end{equation}
The Hubble parameter (and thus the scale factor, up to an irrelevant
scaling) is then fixed by Eq.~\eqref{Hubble}. We seek a two-form
\begin{equation}
\boldsymbol{\sigma}=\sigma(\phi,\dot{\phi})\,\mathrm{d}\dot{\phi}\wedge\mathrm{d}\phi
\end{equation}
that is conserved under evolution, 
\begin{equation}
\pounds_{\mathbf{v}}\boldsymbol{\sigma}=0.\label{preEuler}
\end{equation}
Using the definition of the Lie derivative and rearranging, we can equivalently write in component form
\begin{equation}
\partial_\mu (\sigma v^\mu) = 0,
\label{Euler}
\end{equation}
where $\partial_\mu\equiv\partial/\partial x^\mu$ and $x^\mu = (\phi,\dot{\phi})$. A two-form $\boldsymbol{\sigma}$ for which $\sigma$ satisfies the
Hamiltonian-conservation constraint \eqref{Euler} --- the same as
the Euler equation for stationary fluid flow --- is the natural measure
on the effective phase space, exactly in analogy with the Liouville
measure. We will call the function $\sigma$, which forms the probability
distribution on effective phase space in a given coordinate system,
the \emph{measure density}.

At this point, it is natural to ask whether there is a Lagrangian
description $\mathcal{L}_{\Phi}$ of the trajectories on the effective
phase space $\Phi$. Using Douglas's theorem and the Helmholtz conditions,
we showed \cite{attractors} that there exists a time-independent
Lagrangian description of the equation of motion \eqref{KG} on effective
phase space if and only if there exists a Hamiltonian-conserved measure:
in fact, finding the Lagrangian gives a measure satisfying Eq.~\eqref{Euler}
and vice versa. Further, defining $\pi_{\phi}=\partial\mathcal{L}_{\Phi}/\partial\dot{\phi}$
as the conjugate momentum on $\Phi$, one finds that the Liouville
measure $\mathrm{d}\pi_{\phi}\wedge\mathrm{d}\phi$ on effective phase
space under $\mathcal{L}_{\Phi}$ is just equal to $\sigma\mathrm{d}\dot{\phi}\wedge\mathrm{d}\phi$,
obtained merely by demanding conservation under Hamiltonian evolution.%
\footnote{The corresponding Hamiltonian on effective phase space, ${\mathcal{H}}_{\Phi}=\pi_{\phi}\dot{\phi}-{\mathcal{L}}_{\Phi}$,
is of course not subject to any additional Hamiltonian constraint
as in the full phase space; that is, the Friedmann equation is merely
a redefinition of coordinates on $\Phi$ and does not constrain ${\mathcal{H}}_{\Phi}$.
With this definition, the measure can be written as ${\rm d}{\mathcal{H}}_{\Phi}\wedge{\rm d}t$.%
} For the specific example of $m^{2}\phi^{2}$ inflation, we proved
that such a measure exists and is unique; such an existence/uniqueness
result likely holds for any reasonably well-behaved potential $V\left(\phi\right)$.

\subsection{The space of trajectories\label{trajectoryspace}}

Given the appropriate measure on our effective phase space, one can
use this to determine the natural measure on the space of trajectories.
In general, given some arbitrary measure density on a two-dimensional
manifold and a one-parameter collection of curves that cover the manifold,
there is not a well-defined probability distribution on the set of
curves. However, the Hamiltonian-conserved measure density on effective
phase space is not an arbritary function vis-à-vis the family of trajectories.
Following Refs.~\cite{GHS,attractors}, we can construct a measure
on the space of trajectories in terms of a one-dimensional measure
on any curve transverse to those trajectories, by demanding that the
physical result be independent of our choice of transverse curve.

We begin by choosing some curve in the $\phi$-$\dot{\phi}$ effective
phase space on which to evaluate the measure density $\sigma(\phi,\dot{\phi})$.
For simplicity, we'll imagine choosing $H=$\,constant surfaces,
but any other slicing transverse to the trajectories that evolves
monotonically in time would work just as well.%
\footnote{Note that, regardless of the potential, the scalar equation assures
that $H$ evolves monotonically in time, with $\dot{H}=-\dot{\phi}^{2}/2M_{\mathrm{Pl}}^{2}$.%
} We can reparametrize $\phi$-$\dot{\phi}$ space in terms of $H$
and another coordinate, which we will call $\theta$. For the bundle
of trajectories that, on the $H_{1}$ surface, is centered at $\theta_{1}$
and spans $\mathrm{d}\theta_{1}$, we write the measure as $\left.P(\theta_{1})\right|_{H_{1}}\mathrm{d}\theta_{1}$.
Suppose this bundle of trajectories evolves to $H=H_{2}$, on which
surface it is centered at $\theta_{2}$ and spans $\mathrm{d}\theta_{2}$.
We could equivalently write its probability measure as $\left.P(\theta_{2})\right|_{H_{2}}\mathrm{d}\theta_{2}$.
Of course, the functional forms of $\left.P(\theta_{1})\right|_{H_{1}}$
and $\left.P(\theta_{2})\right|_{H_{2}}$ can be very different.
However, this is the same bundle of trajectories, so for the measure
on the space of trajectories to be well defined, we require 
\begin{equation}
\left.P\left(\theta_{1}\right)\right|_{H_{1}}\mathrm{d}\theta_{1}=\left.P\left(\theta_{2}\right)\right|_{H_{2}}\mathrm{d}\theta_{2}.
\end{equation}

Now, we note that, given a parcel on effective phase space covering
the region $\mathrm{d}\theta_{1}\mathrm{d}H_{1}$ that evolves to
$\mathrm{d}\theta_{2}\mathrm{d}H_{2}$, we have 
\begin{equation}
\sigma\left(H_{1},\theta_{1}\right)\mathrm{d}\theta_{1}\mathrm{d}H_{1}=\sigma\left(H_{2},\theta_{2}\right)\mathrm{d}\theta_{2}\mathrm{d}H_{2}.
\end{equation}
This is just the statement of Liouville's theorem for effective phase
space, \emph{i.e.}, the requirement that $\sigma$ satisfy \eqref{Euler}.
Hence, the correct way to compute $\left.P(\theta)\right|_{H}$, the
probability distribution on the space of trajectories, parametrized
by the coordinate $\theta$ with which the trajectory intersects the
$H$ surface, is 
\begin{equation}
\left.P\left(\theta\right)\right|_{H}\propto\sigma\left(H,\theta\right)\mathrm{d}H.
\end{equation}
We can divide through by $\mathrm{d}t$, since $t$ evolves uniformly
for all trajectories. We therefore have 
\begin{equation}
\left.P\left(\theta\right)\right|_{H}=\frac{\sigma\left(H,\theta\right)|\dot{H}|}{\int\sigma\left(H,\theta^{\prime}\right)|\dot{H}|\mathrm{d}\theta^{\prime}}.\label{trajmeasure}
\end{equation}
Note that we suppressed the arguments $(H,\theta)$ of $\dot{H}$.

Eq.\ \eqref{trajmeasure} is the important expression for this work.
The measure on the space of trajectories is constructed by finding
a conserved measure density $\sigma$ on the effective phase space
and evaluating $|\dot{H}|$ times this measure along a surface of
constant $H$.

As a consistency check, we can derive Eq.~\eqref{trajmeasure} in a slightly
different way. If we had written the effective phase space measure
in the coordinates $(t,\,\theta)$ as $\tilde{\sigma}(t,\theta)\mathrm{d}\theta\wedge\mathrm{d}t=-\sigma(H,\theta)\mathrm{d}\theta\wedge\mathrm{d}H$
(with the minus sign compensating for the fact that $H$ decreases
with $t$, so that $\sigma$ and $\tilde{\sigma}$ are positive) we
could have equivalently defined the measure on the space of trajectories
by explicitly performing the integration over $t$: 
\begin{equation}
\left.P\left(\theta_{0}\right)\right|_{H_{0}}\propto\int_{0}^{\infty}\tilde{\sigma}\left(t,\theta\left(t\right)\right)\mathrm{d}t,
\end{equation}
where the path $(t,\theta(t))$ is chosen such that $\theta(t_{0})=\theta_{0}$
and $H(t_{0})=H_{0}$ for some $t_{0}$. Since $t$ evolves uniformly
for all trajectories, we have 
\begin{equation}
\left.P\left(\theta_{0}\right)\right|_{H_{0}}\propto\left.\tilde{\sigma}\left(t_{0},\theta_{0}\right)\right|_{H(t_{0})=H_{0}}=\sigma\left(H_{0},\theta_{0}\right)|\dot{H}|,
\end{equation}
in agreement with Eq.~\eqref{trajmeasure}.

We are now equipped to make quantitative statements about probabilities
of different FRW trajectories for universes with zero curvature and
compare these predictions for different models of inflation.

\section{The Effective Phase Space Measure for Generic Potentials\label{Generic}}

Before examining specific models of inflation, it will first be informative
to examine the behavior of the effective phase space measure $\boldsymbol{\sigma}$,
without assuming an explicit functional form of the potential, in
two representative classes of inflation: slow roll down a potential
and quasi-de Sitter inflation near a local maximum in a potential,
\emph{i.e.}, a hilltop. The cases are distinct because the fixed point
in effective phase space corresponding to a stationary field at a
potential maximum is a distinguished trajectory by itself and must
be treated carefully.

For this analysis it will be useful to define dimensionless coordinates
\begin{equation}
x=\frac{\phi}{M_{\mathrm{Pl}}}\;\;\;\mathrm{and}\;\;\; y=\frac{\dot{\phi}}{M_{\mathrm{Pl}}^{2}},\label{zcoords0}
\end{equation}
which form a vector
\begin{equation}
\mathbf{x} = (x, y).
\end{equation}
We then define a dimensionless speed in effective phase space 
\begin{equation}
\tilde{\mathbf{v}}\equiv\!\frac{\dot{\mathbf{x}}}{M_{\mathrm{Pl}}}\!=\!\left(y,\!-\tilde{V}^{\prime}(x)-\sqrt{3}y\sqrt{y^{2}/2+\tilde{V}(x)}\right)\!\!,\label{speedz}
\end{equation}
defining $\tilde{V}\left(x\right)\equiv V\left(\phi\left(x\right)\right)/M_{\mathrm{Pl}}^{4}$
as a dimensionless potential and notation $\tilde{V}^{\prime}(x)\equiv\mathrm{d}\tilde{V}/\mathrm{d}x$.

It will also be useful to define a norm for vectors and covectors using a flat fiducial metric: 
\begin{equation}
|\mathbf{c}| \equiv [(c^1)^2 + (c^2)^2]^{1/2}.
\end{equation}
The definition of the norm is simply a mathematical convenience; the fiducial metric should not be regarded as a physical metric on effective phase space. 
It will be convenient in the analysis below, where we derive conditions on the behavior of the measure density, although these conditions would hold even without using the norm notation.
Of course, the physical content of the results is independent of the choice of metric, though the expressions themselves would look different for various choices of norm. As usual, placing bars around scalar quantities, {\it e.g.}, $|\partial_\mu \tilde{v}^\mu|$, simply denotes absolute value.

Define the first potential slow-roll parameter 
\begin{equation}
\epsilon_{V}\equiv\frac{M_{\mathrm{Pl}}^{2}}{2}\left[\frac{V^{\prime}\left(\phi\right)}{V\left(\phi\right)}\right]^{2}=\frac{1}{2}\left[\frac{\tilde{V}^{\prime}\left(x\right)}{\tilde{V}\left(x\right)}\right]^{2}\label{potentialslowroll}
\end{equation}
and the first Hubble slow-roll parameter: 
\begin{equation}
\varepsilon\equiv-\frac{\dot{H}}{H^{2}}=\frac{\dot{\phi}^{2}}{2H^{2}M_{\mathrm{Pl}}^{2}}=3\frac{y^{2}}{y^{2}+2\tilde{V}\left(x\right)}.\label{Hubbleslowroll}
\end{equation}
Substituting Eq.~\eqref{Hubbleslowroll} into Eq.~\eqref{speedz} and rearranging,
one finds 
\begin{equation}
\frac{\left|\tilde{\mathbf{v}}\right|^{2}}{\tilde{V}\left(x\right)^{2}}
=\frac{4\varepsilon^{2}y^{-2}+18\varepsilon}{\left(3-\varepsilon\right)^{2}}+2\epsilon_{V}+6s\frac{\sqrt{2\epsilon_{V}}\sqrt{2\varepsilon}}{3-\varepsilon}.\label{speed-1}
\end{equation}
Here, $s\equiv\mathrm{sgn}[y\tilde{V}^{\prime}(x)]=\pm1$
indicates whether the potential is increasing ($s=+1$) or decreasing
($s=-1$) in the direction along which the field is evolving; we will
generally have $s=-1$ during inflation. Furthermore, after simplifying
with Eq.~\eqref{Hubble}, we have 
\begin{equation}
\frac{\partial_\mu\tilde{v}^\mu}{|y|}=-\frac{3}{\sqrt{2\varepsilon}}-\sqrt{\frac{\varepsilon}{2}},\label{divexact}
\end{equation}
where $\partial_\mu$ denotes partial differentiation with respect to the dimensionless coordinates $x^\mu$ in Eq.~\eqref{zcoords0}. Note that Eqs.~\eqref{speed-1} and \eqref{divexact} are exact expressions:
no slow-roll approximation has yet been made.

\subsection{Slow roll down a potential}

As we discussed in Ref.~\cite{attractors}, slow-roll behavior corresponds
to apparent attractors in effective phase space --- places where the
conserved measure grows large. In this subsection we consider monotonic
slow-roll behavior, characterized by two conditions imposed on Eqs.
\eqref{KG} and \eqref{Hubble}: 
\begin{equation}
\begin{aligned}\dot{\phi}^{2}\ll|V\left(\phi\right)| & \;\;\mathrm{so}\;\; & H^{2}\simeq\frac{1}{3M_{\mathrm{Pl}}^{2}}V\left(\phi\right)\\
|\ddot{\phi}|\ll|H\dot{\phi}|,|V^{\prime}\left(\phi\right)| & \;\;\mathrm{so}\;\; & 3H\dot{\phi}\simeq-V^{\prime}\left(\phi\right).
\end{aligned}
\label{slowrollassumptions}
\end{equation}
Then $\varepsilon\approx\epsilon_{V}\equiv\epsilon\ll1$ and we have
from Eq.~\eqref{speed-1}: 
\begin{equation}
\frac{\left|\tilde{\mathbf{v}}\right|^{2}}{\tilde{V}\left(x\right)^{2}}\simeq\frac{4\epsilon^{2}}{9y^{2}}+4(1+s)\epsilon.
\end{equation}
Further, imposing $H^{2}\ll M_{\mathrm{Pl}}^{2}$, so $\varepsilon\gg y^{2}$,
\begin{equation}
\left|\tilde{\mathbf{v}}\right|\simeq|y|\simeq\sqrt{\frac{2}{3}\epsilon\tilde{V}(x)}.\label{speedSR}
\end{equation}
Similarly, in the slow-roll regime, 
\begin{equation}
\partial_\mu\tilde{v}^\mu\simeq-\frac{3|y|}{\sqrt{2\epsilon}}\simeq-\sqrt{3\tilde{V}(x)}.\label{divSR}
\end{equation}
Note that the second slow-roll condition in Eq.~\eqref{slowrollassumptions}
does not necessarily apply near a hilltop, as $|H\dot{\phi}|\gg|\ddot{\phi}|$
can fail. This is an important distinction; as we will see, behavior
of the measure density near a hilltop in effective phase space is
very different from what we find for trajectories that are uniformly
slowly rolling down a potential. Our slow-roll conditions in this
subsection are most compatible with potentials with $V^{\prime\prime}(\phi)>0$,
such as monomial models, in which hilltop behavior is manifestly absent.

Now, we examine what implications our analysis has for the form of
the measure $\boldsymbol{\sigma}=\sigma(x,y)\mathrm{d}y\wedge\mathrm{d}x$
on effective phase space. With the requirement \eqref{Euler} that the measure be conserved under Hamiltonian evolution, Eqs.~\eqref{speedSR} and \eqref{divSR} imply that
\begin{equation}
\frac{(\partial_\mu \ln\sigma)\tilde{v}^\mu}{|\tilde{\mathbf{v}}|}=\frac{|\partial_\mu \tilde{v}^\mu|}{|\tilde{\mathbf{v}}|}\simeq\frac{3}{\sqrt{2\epsilon}}\label{gradient}
\end{equation}
near the slow-roll regime and for $H^{2}\ll M_{\mathrm{Pl}}^{2}$.
Note that the left side of Eq.~\eqref{gradient} is just the gradient of $\ln\sigma$ along a slow-roll curve; thus, the closer to slow roll we approach and the farther along a slow-roll trajectory we progress, the larger $\sigma$ becomes. In particular, for a slow-roll trajectory that evolves from $\mathbf{x}_1$ to $\mathbf{x}_2$ in effective phase space, we have
\begin{equation}
\frac{\sigma(\mathbf{x}_2)}{\sigma(\mathbf{x}_1)} \simeq \exp\left[3\int_{\mathrm{C}} \frac{\mathrm{d}\ell}{\sqrt{2\epsilon(\mathbf{x})}}\right],\label{measurebehaviorslowroll}
\end{equation}
where $\mathrm{C}$ is the segment of the slow-roll curve in the plane between $\mathbf{x}_1$ and $\mathbf{x}_2$ and $\mathrm{d}\ell$ is the line element along this curve, defined with respect to the fiducial metric.
Hence, we generically expect that any region in effective phase space
satisfying our slow-roll conditions \eqref{slowrollassumptions} will
have large measure density $\sigma$ on effective phase space, relative
to nearby regions. Of course, the measure density on effective phase
space can be large in regions that fail the slow-roll conditions,
such as during reheating, in which apparent attractor solutions traverse
long paths in compact regions of effective phase space and trajectories
appear to converge. All of these statements are made with regard to
the phase space measure, not the measure on the space of trajectories;
the factor of $|\dot{H}|$ in Eq.~\eqref{trajmeasure} makes this
an important distinction.

\subsection{Inflation on a hilltop potential\label{hilltop}}

In a model of inflation governed by a potential with a hilltop (a
local maximum), there are two types of classical solutions that differ
qualitatively from the usual picture of the inflaton field rolling
down the potential and reheating: 1) fixed point trajectories, \emph{i.e.},
exactly de Sitter solutions, which start with $\dot{\phi}=0$ at the
top of the potential and inflate forever; and 2) roll-up trajectories,
which start from somewhere on the slope of the potential and asymptotically
approach the fixed point. The \emph{fixed point} is the location $(\phi,\dot{\phi}=0$),
equivalently $(x,y=0)\equiv\mathbf{x}_{0}$ in effective phase
space, for which $\phi$ is at the hilltop of $V(\phi)$. We would
like to elucidate the behavior of the phase space measure density
$\sigma(x,y)$ in the region of effective phase space near
the fixed point.

Near the fixed point, $\dot{\phi}^{2}\ll V(\phi)$, so Eq.~\eqref{divexact}
implies $|\partial_\mu\tilde{v}^\mu|\rightarrow(3\tilde{V}(x))^{1/2}$,
as in Eq.~\eqref{divSR}; moreover, $\tilde{\mathbf{v}}\rightarrow0$.
With Eq.~\eqref{Euler} requiring conservation of the measure under Hamiltonian evolution, the Cauchy--Schwarz inequality implies:
\begin{equation}
|\partial_\mu\tilde{v}^\mu|\sigma\leq|\boldsymbol{\partial}\sigma||\tilde{\mathbf{v}}|,
\end{equation}
where we use the vector notation $\boldsymbol{\partial}$ for $\partial_\mu$. Since $|\partial_\mu\tilde{v}^\mu|$ is finite and $|\tilde{\mathbf{v}}|\rightarrow0$,
requiring that $\sigma$ be smooth implies 
\begin{equation}
\sigma(\mathbf{x})\rightarrow0\;\;\mathrm{as}\;\;\mathbf{x}\rightarrow\mathbf{x}_{0}.
\end{equation}
Even if we relaxed the assumption of regularity, we can still show
that $\sigma$ is small near the fixed point. We observe, given a
smooth, slowly-varying potential $V(\phi)$, that any fixed point
in effective phase space will be at the terminus of an apparent attractor,
that is, a region where both slow-roll conditions \eqref{slowrollassumptions}
are met. As trajectories flow from near the fixed point along the
apparent attractor, the first condition is always met, while the second
becomes an increasingly good approximation. Hence, our conclusion
from the slow-roll regime becomes applicable and so Eq.~\eqref{measurebehaviorslowroll}
implies that the effective phase space measure near the fixed point
is exponentially suppressed compared with the measure further down
the slow-roll apparent attractor. Other than the fixed point trajectory
itself --- which is irrelevant to inflation, since the field does
not evolve --- there is, relative to the slow-roll regime, very little
measure near the hilltop. Recall that slicing effective phase space
into sets of constant $H$ to parametrize the space of trajectories
incurs an additional factor of $|\dot{H}|$ to convert the phase space
measure density into the probability distribution on the trajectories;
this suppresses the measure assigned to the roll-up trajectories even
more. However, we have shown here that roll-up trajectories are suppressed
in the canonical measure on effective phase space, even without the
help of this additional factor. Since the measure is conserved under
Hamiltonian evolution, any roll-up trajectory, \emph{i.e.}, the FRW
evolution that comes arbitrarily close to de Sitter, is a set of measure
zero.

\section{Quadratic Inflation}

\label{Quadratic}

\subsection{Preliminaries}

As a representative example of slow-roll inflation with $V^{\prime\prime}(\phi)>0$,
we consider monomial inflation with a quadratic potential, 
\begin{equation}
V(\phi)=\frac{1}{2}m^{2}\phi^{2}.\label{potential1}
\end{equation}
If the recent BICEP2 discovery of $B$-mode polarization \cite{Bicep2}
is the result of primordial gravitational waves, then this simple
model is in good agreement with the observed tensor perturbations.
A canonical model in the inflationary literature \cite{Lindequad,chaotic}
and one of theoretical interest \cite{Kawasaki}, the set of quadratic
and related potentials is an important area of current investigation
\cite{Ellisquad1,axionquad}, given the status of observations
\cite{Bicep2,Planck}.

It will eventually be useful to redefine our dimensionless coordinates $\mathbf{x}$
differently from those in Eq.~\eqref{zcoords0}: 
\begin{equation}
x=\frac{\phi}{\sqrt{6}M_{\mathrm{Pl}}}\;\;\;\mathrm{and}\;\;\; y=\frac{\dot{\phi}}{\sqrt{6}mM_{\mathrm{Pl}}}.\label{zcoords1}
\end{equation}
We define polar coordinates $(z,\theta)$:
\begin{equation}
z\equiv\sqrt{x^2 + y^2}=\frac{H}{m}\label{zcoords2}
\end{equation}
and
\begin{equation}
\tan\theta=\frac{y}{x}=\frac{\dot{\phi}}{m\phi},\label{theta}
\end{equation}
so $\dot{\phi}=\sqrt{6}mM_{\mathrm{Pl}}z\sin\theta$ and $\phi=\sqrt{6}M_{\mathrm{Pl}}z\cos\theta$.

\begin{figure}
\noindent \begin{centering}
\includegraphics[height=2in]{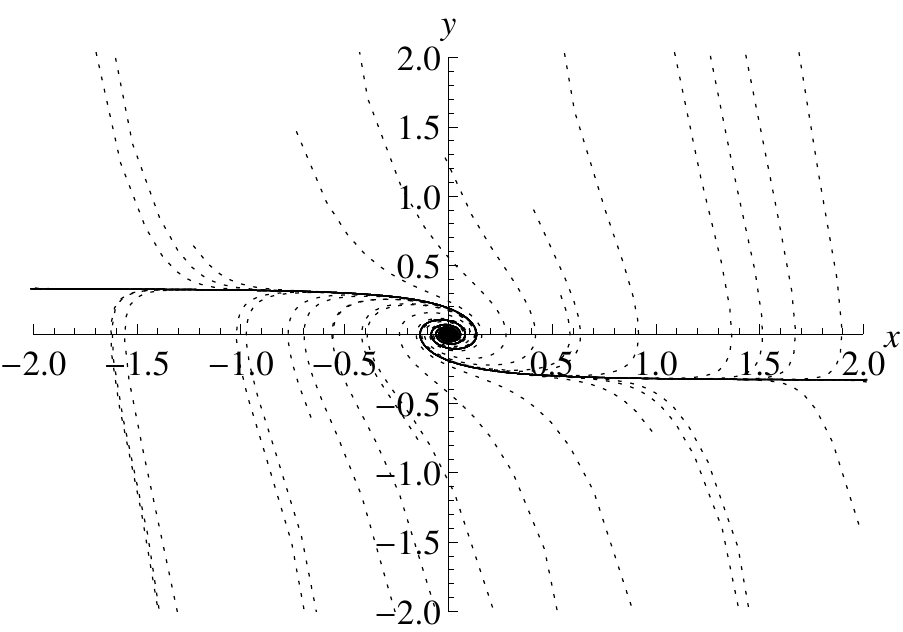} 
\par\end{centering}

\caption{\label{quadraticattractors} Trajectories in effective phase space
for quadratic inflation. The field value and velocity are parametrized
by the variables $x$ and $y$, defined in Eq.~\eqref{zcoords1}.
The dark nearly-horizontal lines indicate the apparent attractors,
where the conserved measure grows large. For clarity we used the (unrealistic)
value of $m=0.2M_{\mathrm{Pl}}$ to make this plot.}
\end{figure}

Using Eq.~\eqref{KG}, we can plot trajectories in the $\phi$-$\dot{\phi}$
plane and see explicitly the effective phase space behavior, as shown
in Fig.~\ref{quadraticattractors}. In particular, note the apparent
attractor solutions that appear at $y=\pm1/3$, corresponding
to $\dot{\phi}=\pm\sqrt{2/3}mM_{\mathrm{Pl}}$. Of course, in a strict
phase-space sense, these ``attractors'' are illusory \cite{attractors}.
In the Liouville measure, phase-space density is conserved. The apparent
attractor behavior actually indicates that the measure density grows
large in that region.

The apparent attractor solution at $\dot{\phi}=\pm\sqrt{2/3}mM_{\mathrm{Pl}}$
intersects the $H=$\,constant ellipse at 
\begin{equation}
|\sin\theta|=\frac{m}{3H}.\label{attrangle}
\end{equation}
In the early universe ($H\gg m$), we therefore have $\theta\simeq0$
or $\pi$ on the apparent attractor.

\subsection{Counting $e$-folds}

For the quadratic potential \eqref{potential1}, one has the potential
slow-roll parameter from Eq.~\eqref{potentialslowroll}: 
\begin{equation}
\epsilon_{V}=2\left(\frac{M_{\mathrm{Pl}}}{\phi}\right)^{2}.
\end{equation}
Inflation (and counting of $e$-folds $N$) ends when $\epsilon_{V}=1$,
which occurs at $\phi_{\mathrm{f}}=\sqrt{2}M_{\mathrm{Pl}}$.

For slow roll, $H^{2}\simeq V/3M_{\mathrm{Pl}}^{2}$, the scalar equation
\eqref{KG} becomes $3H\dot{\phi}\simeq-V^{\prime}$ and so $H\mathrm{d}t\simeq\pm\mathrm{d}\phi/\sqrt{2\epsilon_{V}}M_{\mathrm{Pl}}$.
Thus, when the field value is $\phi$, the number of $e$-folds remaining
before the end of inflation is 
\begin{equation}
N\left(\phi\right)=\left|\int_{|\phi|}^{\sqrt{2}M_{\mathrm{Pl}}}\frac{1}{\sqrt{2\epsilon_{V}}}\frac{\mathrm{d}\phi^{\prime}}{M_{\mathrm{Pl}}}\right|=\frac{1}{4}\left(\frac{\phi}{M_{\mathrm{Pl}}}\right)^{2}-\frac{1}{2},\label{Nphi}
\end{equation}
which is accurate as long as the slow-roll conditions \eqref{slowrollassumptions} are satisfied. While exact number of $e$-folds, defined in Eq.~\eqref{efoldsdefined} using the full expression for $H\mathrm{d}t$ given in Eq.~\eqref{Hubble}, would have corrections near the end of inflation where the slow-roll conditions begin to break down, we shall see that this will not appreciably affect the total number of $e$-folds that we ultimately compute.
Consider a trajectory that starts at angle $\theta$ on the surface
where $H=M_{\mathrm{Pl}}$. We will call this the Planck surface;
of course, one could choose a different ultraviolet cutoff $\Lambda_{\mathrm{UV}}\gg m$
for the effective field theory, on which to start evaluating trajectories
at time $t=0$. In that case, one could simply replace $M_{\mathrm{Pl}}$
by $\Lambda_{\mathrm{UV}}$ as appropriate in all of our $e$-fold
counting. For simplicity we will choose $\Lambda_{\mathrm{UV}}=M_{\mathrm{Pl}}$.
The initial field value for $\phi$ is then $\sqrt{6}(\cos\theta)M_{\mathrm{Pl}}^{2}/m$.

In the $H\gg m$ region of $\phi$-$\dot{\phi}$ space, trajectories
snap quickly to the apparent attractor, with $\dot{\phi}$ changing
much faster than $m\phi$. That is, using the scalar equation, we have
\begin{equation}
\frac{\dot{\mathbf{x}}}{m}=\left(y,-x-3y\sqrt{x^{2}+y^{2}}\right).\label{dotz}
\end{equation}
Thus, when $z=\sqrt{x^{2}+y^{2}}\gg1$, we have $\dot{y}\gg\dot{x}$,
as claimed. Hence, $x(t=0)$, to a very good approximation, is
equal to $x$ at the time when the trajectory starts the slow-roll
process. Therefore, we can write the \emph{total} number of $e$-folds
that this trajectory (parametrized by $\theta$ on the Planck surface)
undergoes as 
\begin{equation}
N_{\mathrm{tot}}=\frac{3}{2}\left(\frac{M_{\mathrm{Pl}}}{m}\right)^{2}\cos^{2}\theta-\frac{1}{2}\simeq\frac{3}{2}\left(\frac{M_{\mathrm{Pl}}}{m}\right)^{2}\cos^{2}\theta.\label{Nanalytic}
\end{equation}
Maximal inflation occurs when $\theta\simeq0$ or $\pi$, \emph{i.e.},
the trajectory starts out near the apparent attractor at the Planck
scale, which gives 
\begin{equation}
N_{\mathrm{max}}=\frac{3}{2}\left(\frac{M_{\mathrm{Pl}}}{m}\right)^{2}\label{Nmax}
\end{equation}
$e$-folds of inflation. Comparing the analytical prediction \eqref{Nanalytic}
with numerical simulation, we find very good agreement.

\subsection{How many $e$-folds should we expect in quadratic inflation?\label{HowManyQuadratic}}

We know from Eq.~\eqref{Nanalytic} how to predict the total number
of $e$-folds of inflation a trajectory will undergo based on a particular
parametrization of the family of trajectories, namely, by the angular
coordinate $\theta$ with which the trajectory intersects a surface
of particular energy density, in this case $H=M_{\mathrm{Pl}}$. We
would now like to ask the question of how many $e$-folds we should
expect, using the prescription for finding the appropriate measure
\eqref{trajmeasure} on the space of trajectories, as described in
Sec.~\ref{trajectoryspace}.

First, we need to find the measure $\boldsymbol{\sigma}$ on effective
phase space for quadratic inflation. In the $(z,\theta)$ coordinates
defined in Eqs.~\eqref{zcoords2} and \eqref{theta}, we can write
the velocity \eqref{velocity} of trajectories in effective phase
space as 
\begin{equation}
\mathbf{v}=\dot{\mathbf{x}}=-3mz^{2}\sin^{2}\theta\mathbf{\hat{z}}-m\left(z+3z^{2}\sin\theta\cos\theta\right)\boldsymbol{\hat{\theta}},
\end{equation}
where 
\begin{equation}
\left(\begin{array}{c}
\mathbf{\hat{z}}\\
\boldsymbol{\hat{\theta}}
\end{array}\right)=\left(\begin{array}{cc}
\cos\theta & \sin\theta\\
-\sin\theta & \cos\theta
\end{array}\right)\left(\begin{array}{c}
\mathbf{\hat{x}}\\
\mathbf{\hat{y}}
\end{array}\right).\label{coordrotation}
\end{equation}

In the early universe, when $H\gg m$ (such as on the Planck ellipse
$H=M_{\mathrm{Pl}}$), we have $z\gg1$. Thus, $\mathbf{v}$ becomes
approximately 
\begin{equation}
\mathbf{v}\simeq-3mz^{2}\sin^{2}\theta\mathbf{\hat{z}}-3mz^{2}\sin\theta\cos\theta\boldsymbol{\hat{\theta}},\label{highzspeed}
\end{equation}
and so the requirement \eqref{Euler} for $\boldsymbol{\sigma}$ to
be conserved under Hamiltonian evolution becomes 
\begin{equation}
\partial_{\theta}\sigma=-z\tan\theta\partial_{z}\sigma-\left(2\tan\theta+\cot\theta\right)\sigma.
\end{equation}
The general solutions for $\sigma$ take the form \cite{attractors}
\begin{equation}
\sigma=\sum_{\gamma}C_{\gamma}z^{\gamma-3}\left|\frac{\cos^{\gamma-1}\theta}{\sin\theta}\right|,\label{measurezlarge}
\end{equation}
for $z\gg1$, where $\gamma,\, C_{\gamma}\in\mathbb{R}$. We note
that $\sigma$ diverges along the $\sin\theta=0$ axis, corresponding
to the buildup of trajectories along the apparent attractor; in an
exact numerical solution, the distribution $\sigma$ would become
large on the apparent attractor solution, as is clear from Fig.~\ref{quadraticattractors},
using the fluid flow analogy. For the potential \eqref{potential1},
we proved in Ref.~\cite{attractors} that the measure $\boldsymbol{\sigma}$
has a unique solution; hence, many possible solutions in Eq.~\eqref{measurezlarge}
are spurious or unphysical.

As we can see from flow of the vector field shown in Fig.~\ref{quadraticattractors},
we should require that $\sigma$ be finite everywhere except on the
apparent attractor solution; imposing this condition requires $\gamma\geq1$.
Further, at fixed $\theta$, trajectories become more squeezed together
as $z$ decreases, since more and more time evolution is compressed
into a smaller and smaller range of $H$. Hence, we should require
$\sigma$ to be a non-increasing function of $z$ at fixed $\theta$,
so $\gamma\leq3$. Imposing the further requirement that $\sigma$
be infinitely differentiable everywhere except the apparent attractor
solution selects $\gamma=3$ as the physical solution, so we end up
with 
\begin{equation}
\sigma(H=M_{\mathrm{Pl}},\theta)\propto\left|\frac{\cos^{2}\theta}{\sin\theta}\right|.\label{sigmaPlanck}
\end{equation}

In Sec.~\ref{trajectoryspace} we demonstrated how to obtain the measure
on the space of trajectories from $\boldsymbol{\sigma}$, \emph{cf.}
Ref.~\cite{GHS}. Specifically, the probability distribution on the
space of trajectories, parametrized by $\theta$ on some surface of
constant $H$, is, up to normalization, given by $\sigma\left(H,\theta\right)|\dot{H}|$.
Using Eq.~\eqref{KG}, we have in the $(z,\theta)$ coordinates: 
\begin{equation}
\dot{H}=-\frac{\dot{\phi}^{2}}{2M_{\mathrm{Pl}}^{2}}=-3m^{2}z^{2}\sin^{2}\theta.\label{Hdot}
\end{equation}
Thus, the probability distribution on the space of trajectories on
the Planck surface (where $z=M_{\mathrm{Pl}}/m=$\,constant), parametrized
by the coordinate $\theta$, is 
\begin{equation}
\left.P\left(\theta\right)\right|_{H=M_{\mathrm{Pl}}}=\frac{3}{4}\left|\cos^{2}\theta\sin\theta\right|.\label{trajmeasure-1}
\end{equation}
The overall normalization has been fixed by requiring $\int\mathrm{d}\theta\, P(\theta)=1$.

Finally, we can now compute the expected total number of $e$-folds
of inflation, using the canonical measure \eqref{trajmeasure-1} and
our $e$-fold counting \eqref{Nanalytic}: 
\begin{equation}
\begin{aligned}\left\langle N_{\mathrm{tot}}\right\rangle  & =\int_{0}^{2\pi}N\left(\theta\right)\left.P\left(\theta\right)\right|_{H=M_{\mathrm{Pl}}}\mathrm{d}\theta\\
 & =\frac{9}{8}\left(\frac{M_{\mathrm{Pl}}}{m}\right)^{2}\int_{0}^{2\pi}\cos^{4}\theta\left|\sin\theta\right|\mathrm{d}\theta\\
 & =\frac{9}{10}\left(\frac{M_{\mathrm{Pl}}}{m}\right)^{2}\\
 & =\frac{3}{5}N_{\mathrm{max}}.
\end{aligned}
\label{Ntotal1}
\end{equation}

Now, assuming a quadratic potential \eqref{potential1}, the amplitude
of observed CMB scalar perturbations is 
\begin{equation}
\Delta_{\mathrm{s}}^{2}\left(k_{\mathrm{CMB}}\right)=\frac{1}{6\pi^{2}}\left(\frac{m}{M_{\mathrm{Pl}}}\right)^{2}N_{\mathrm{CMB}}^{2},
\end{equation}
where $N_{\mathrm{CMB}}\approx50$ is the number of $e$-folds between
horizon exit of CMB scales and the end of inflation. Using the Planck
observations \cite{Planck} for the amplitude of scalar perturbations,
we have $m=7\times10^{-6}M_{\mathrm{Pl}}=2\times10^{13}$\,GeV, which
implies that for quadratic inflation we expect 
\begin{equation}
\left\langle N_{\mathrm{tot}}\right\rangle =2\times10^{10}.
\end{equation}
That is, typical universes under the canonical measure \eqref{trajmeasure-1}
with the inflaton mass we obtain by positing a quadratic potential
\eqref{potential1} and requiring consistency with Planck \cite{Planck}
undergo much more than the required number of $e$-folds needed to
solve the horizon problem; hence, one can view our observed Universe
as natural in the theory of quadratic inflation, with regard to the
canonical measure (with the caveats about inhomogeneities noted in
the Introduction).

Looking at the conclusion another way, we note that the probability
for $N_{\mathrm{tot}}$ to be greater than some particular value $N_{0}$
is just the probability that $\cos^{2}\theta>(2/3)(m/M_{\mathrm{Pl}})^{2}N_{0}\equiv\cos^{2}\zeta$.
Thus, 
\begin{equation}
\begin{aligned}\mathrm{Pr}\left(N_{\mathrm{tot}}>N_{0}\right) & =4\times\frac{3}{4}\int_{0}^{\zeta}\cos^{2}\theta\sin\theta\mathrm{d}\theta\\
 & =1-\left(\frac{2}{3}\right)^{3/2}\left(\frac{m}{M_{\mathrm{Pl}}}\right)^{3}N_{0}^{3/2}\\
 & =1-\left(\frac{N_{0}}{N_{\mathrm{max}}}\right)^{3/2},
\end{aligned}
\label{probquad}
\end{equation}
where $N_{\mathrm{max}}$ is defined in Eq.~\eqref{Nmax}. That is,
if $m=2\times10^{13}$\,GeV, the probability of having fewer than
50 $e$-folds of inflation is of order $10^{-13}$. Differentiating
Eq.~\eqref{probquad}, the measure on the space of trajectories can
be written in terms of the number $N_{\mathrm{tot}}$ of $e$-folds
ultimately achieved, between zero and $N_{\mathrm{max}}$: 
\begin{equation}
P\left(N_{\mathrm{tot}}\right)\mathrm{d}N_{\mathrm{tot}}=\frac{3}{2N_{\mathrm{max}}^{3/2}}N_{\mathrm{tot}}^{1/2}\mathrm{d}N_{\mathrm{tot}}.
\end{equation}
Universes that undergo 50 or more $e$-folds of inflation, like our
own, are overwhelmingly generic from the perspective of the canonical
measure for high-scale quadratic inflation.

The specific number $\left\langle N_{\mathrm{tot}}\right\rangle =2\times10^{10}$
is suggestive, but it shouldn't be taken too literally. In quadratic
inflation, the field has a value $\phi\sim10M_{\mathrm{Pl}}$ at the
epoch when currently observable large-scale perturbations are being
generated; our calculation fearlessly extrapolates the functional
form of the potential to values of order $10^{5}M_{\mathrm{Pl}}$,
where there is little reason for it to be trusted. Nevertheless, we
expect that our result has a robust physical interpretation for more
general potentials: in large-field inflation, when the potential increases
to the Planck limit, it is natural to achieve a large amount of inflation.
There are certainly some trajectories that spend little or no time
on the apparent attractor, remaining dominated by kinetic energy all
the way up to Planck densities. Our results suggest that, in large-field
inflation, such trajectories are extremely unlikely, as generic evolution
quickly snaps to the apparent attractor, yielding many $e$-folds
of inflation.

\section{Cosine (Natural) Inflation}

\label{Cosine}

\subsection{Preliminaries}

We now turn to the model of cosine or ``natural'' inflation \cite{Natural1,Natural2},
in which the inflaton $\phi$ could be a pseudo-Nambu--Goldstone boson
$\theta=\phi/f$ with a global shift symmetry broken at scale $f$.
The global symmetry is explicitly broken at scale $\Lambda$, giving
the boson a mass, via a potential 
\begin{equation}
V\left(\phi\right)=\Lambda^{4}\left[1-\cos\left(\phi/f\right)\right].\label{potential2}
\end{equation}
Cosine inflation is representative of the general class of hilltop
inflation models: the inflaton potential has a region where $V^{\prime\prime}(\phi)<0$.
Qualitatively, this model has similarities and differences with monomial
inflation models. Like monomial inflation, it can exhibit slow-roll
behavior. Unlike monomial models, however, hilltop models have trajectories
in which the inflaton stays near the top of the potential for a parametrically
long time and pure de Sitter space is allowed if the field sits exactly
at the potential maximum. Such trajectories would seem to allow hilltop
models to achieve a very large number of $e$-foldings without the
concomitant large excursion in field values endemic to monomial models
and potentially troublesome from the effective field theory perspective.
Ultraviolet completions of cosine inflation models have been investigated
\cite{Extranatural,naturalUV,Kallosh}, which improve the applicability
of effective field theoretic reasoning. Cosine inflation models are
of significant current interest \cite{Freese} and generically have
regions of parameter space that can achieve agreement with observations
from BICEP2 \cite{Bicep2} and Planck \cite{Planck}. In cosine inflation,
the field can without loss of generality be restricted to the interval
between $\pm\pi f$, with the periodic identification $\phi\sim\phi+2\pi f$
as an equivalence class.

As for quadratic inflation, we will find dimensionless coordinates
useful (different from Eqs.~\eqref{zcoords0} and \eqref{zcoords1}):
\begin{equation}
x=\sqrt{\frac{2}{3}}\frac{f}{M_{\mathrm{Pl}}}\sin\left(\phi/2f\right)\;\;\;\mathrm{and}\;\;\; y=\frac{f\dot{\phi}}{\sqrt{6}\Lambda^{2}M_{\mathrm{Pl}}}.\label{coords}
\end{equation}
Because of the restricted range of the field, $x$ is isomorphic
to $\phi$, so our discussion about vector field invariance from Ref.
\cite{attractors} applies and $(x,y)$ forms an effective
phase space. Note that $x\in[-\sqrt{2/3}f/M_{\mathrm{Pl}},\sqrt{2/3}f/M_{\mathrm{Pl}}]$,
with the identification $x\sim x+2\sqrt{2/3}f/M_{\mathrm{Pl}}$.
As before, define polar coordinates 
\begin{equation}
z\equiv\sqrt{x^{2}+y^{2}}=\frac{f}{\Lambda^{2}}H
\end{equation}
and 
\begin{equation}
\tan\theta=\frac{y}{x}=\frac{\dot{\phi}}{2\Lambda^{2}\sin\left(\phi/2f\right)}.
\end{equation}
Because $x$ can only take values between $\pm\sqrt{2/3}f/M_{\mathrm{Pl}}$,
the Planck surface $H=M_{\mathrm{Pl}}$ subtends a finite set of angles
$[\theta_{0},\pi-\theta_{0}]\cup[\pi+\theta_{0},2\pi-\theta_{0}]$,
where 
\begin{equation}
\cos\theta_{0}=\frac{\sqrt{\frac{2}{3}}\frac{f}{M_{\mathrm{Pl}}}}{\frac{f}{\Lambda^{2}}M_{\mathrm{Pl}}}=\sqrt{\frac{2}{3}}\frac{\Lambda^{2}}{M_{\mathrm{Pl}}^{2}}\ll1,\label{x}
\end{equation}
\emph{i.e.}, $\theta_{0}$ is close to $\pi/2$ or $3\pi/2$.

\begin{figure}
\noindent \begin{centering}
\includegraphics[height=2in]{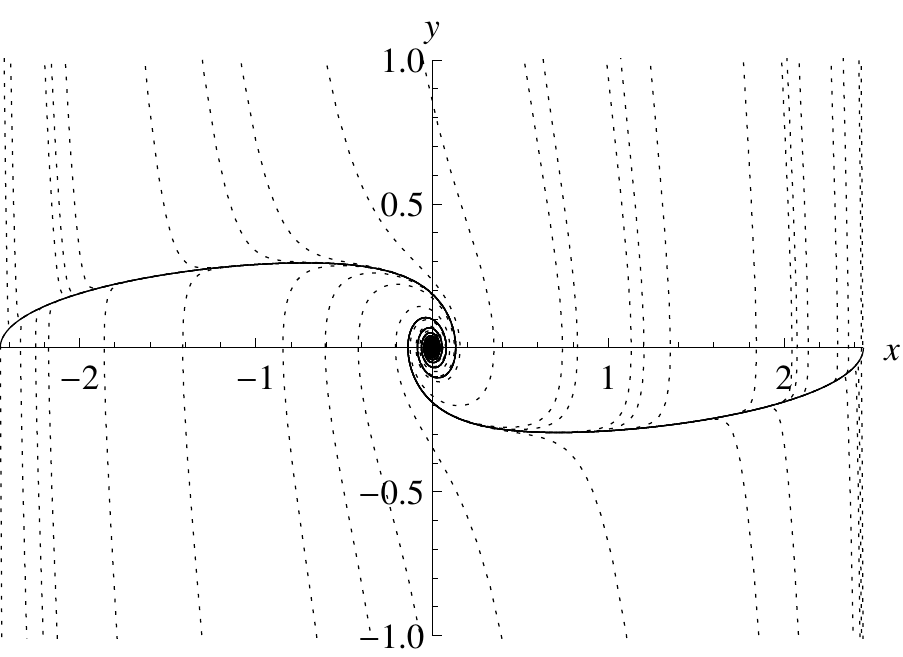} 
\par\end{centering}

\caption{\label{cosineattractors} Trajectories in effective phase space for
cosine inflation. The field value and velocity are parametrized by
the variables $x$ and $y$, defined in Eq.~\eqref{coords}.
The dark spirals indicate the apparent attractors, where the conserved
measure grows large. For this plot we used $f=3M_{\mathrm{Pl}}$,
$\Lambda=0.1M_{\mathrm{Pl}}$.}
\end{figure}

In $(x,y)$ coordinates, the velocity vector $\mathbf{v}=\dot{\mathbf{x}}$,
using Eq.~\eqref{KG}, is 
\begin{equation}
\begin{aligned}\frac{f\mathbf{v}}{\Lambda^{2}} & =y\sqrt{1-\frac{3}{2}\frac{M_{\mathrm{Pl}}^{2}}{f^{2}}x^{2}}\,\,\mathbf{\hat{x}}\\
 & \hspace{1em}-\Bigg(3y\sqrt{x^{2}+y^{2}}
+x\sqrt{1-\frac{3}{2}\frac{M_{\mathrm{Pl}}^{2}}{f^{2}}x^{2}}\Bigg)\mathbf{\hat{y}},
\end{aligned}
\label{speed}
\end{equation}
or equivalently, in polar coordinates $(z,\theta)$,
\begin{equation}
\begin{aligned}
\frac{f\mathbf{v}}{\Lambda^{2}} &=-3z^{2}\sin^{2}\theta\,\,\mathbf{\hat{z}}\\
 & \hspace{1em}-\Bigg(\!3z^{2}\sin\theta\!\cos\theta 
 \!+\!z\sqrt{1\!-\!\frac{3}{2}\frac{M_{\mathrm{Pl}}^{2}}{f^{2}}z^{2}\!\cos^{2}\theta}\Bigg)\boldsymbol{\hat{\theta}},
\end{aligned}
\label{speed2}
\end{equation}
where $\mathbf{\hat{x}}$, $\mathbf{\hat{y}}$, $\mathbf{\hat{z}}$,
and $\boldsymbol{\hat{\theta}}$ are related as in Eq.~\eqref{coordrotation}.
Plotting integral curves of this vector field, one can visualize trajectories
in effective phase space, shown in Fig.~\ref{cosineattractors}.
As for quadratic inflation, there is an apparent attractor, but for
$\phi$ near $\pm f$, lingering behavior near the hilltop is also
possible.

\subsection{Counting $e$-folds}

With the potential slow-roll parameter $\epsilon_{V}$ defined as
in Eq.~\eqref{potentialslowroll}, for the cosine inflation potential
\eqref{potential2} one has 
\begin{equation}
\epsilon_{V}=\frac{M_{\mathrm{Pl}}^{2}}{2f^{2}}\frac{\sin^{2}\left(\phi/f\right)}{\left[1-\cos\left(\phi/f\right)\right]^{2}}=\frac{1-b^{2}x^{2}}{3x^{2}},\label{potentialslownatural}
\end{equation}
for convenience defining a constant 
\begin{equation}
b\equiv\sqrt{3/2}M_{\mathrm{Pl}}/f.\label{b}
\end{equation}
Inflation --- and counting of $e$-folds --- ends when $\epsilon_{V}=1$,
which occurs at $|x|=\left(3+b^{2}\right)^{-1/2}$.

For slow roll and assuming $\ddot{\phi}$ is small compared to other
terms in the scalar equation \eqref{KG}, we have $H^{2}\simeq V/3M_{\mathrm{Pl}}^{2}$
and $3H\dot{\phi}\simeq-V^{\prime}$, so 
\begin{equation}
H\mathrm{d}t\simeq\pm\frac{\mathrm{d}\phi}{\sqrt{2\epsilon_{V}}M_{\mathrm{Pl}}}=\pm\frac{3|x|\mathrm{d}x}{1-b^{2}x^{2}},\label{Hdtnatural}
\end{equation}
after using Eqs.~\eqref{coords} and \eqref{potentialslownatural}.%
\footnote{Though in general a hilltop trajectory can violate the condition that
$3H\dot{\phi}\simeq-V^{\prime}$, one can show that, for the potential
\eqref{potential2}, the total number of $e$-folds we compute is
accurate even without this assumption. See footnote \ref{approx}.%
} Thus, when the field value is $x$, the number of $e$-folds
remaining before the end of inflation is 
\begin{equation}
\begin{aligned}N\left(x\right) & =\left|\int_{|x|}^{\left(3+b^{2}\right)^{-1/2}}\frac{3x^{\prime}\mathrm{d}x^{\prime}}{1-b^2 (x^{\prime})^2}\right|\\
 & =\frac{3}{2b^{2}}\ln \left[\frac{1}{\left(1-b^{2}x^{2}\right)\left(1+\frac{1}{3}b^{2}\right)}\right].
\end{aligned}
\label{Nz1}
\end{equation}

We would like to parametrize the number of $e$-folds a trajectory
undergoes based upon its coordinate $\theta$ on the Planck surface,
\emph{not} its coordinate $x$ when it enters the slow-roll regime.
From the vector field in Eqs.~\eqref{speed} and \eqref{speed2}, we see that when $z=\sqrt{x^{2}+y^{2}}\gg1$
and $y\gg x$ (which is true on the Planck surface) we have
$\dot{y}\gg\dot{x}$. Therefore, as for quadratic inflation,
we are able to approximate $x\left(\mathrm{Planck\; surface}\right)\simeq x\left(\mathrm{enter\; slow\; roll}\right)$
for a given trajectory.%
\footnote{Note that this approximation leads us to assign nonzero measure to
the set of trajectories that come arbitrarily close to the fixed point.
It therefore assigns nonzero measure where the roll-up trajectory
intersects the Planck surface, which we argued in Sec.~\ref{Generic}
is not strictly correct. However, if anything, this assumption should
overestimate the expected total number of $e$-folds.%
} The total number of $e$-folds attained by a trajectory that starts
out at angle $\theta$ on the Planck surface is then 
\begin{equation}
N_{\mathrm{tot}}\left(\theta\right)=\frac{3}{2b^{2}}\ln\left[\frac{1}{\left(1-\frac{\cos^{2}\theta}{\cos^{2}\theta_{0}}\right)\left(1+\frac{1}{3}b^{2}\right)}\right],\label{Nanalytic2}
\end{equation}
where $\theta_{0}$ is defined in Eq.~\eqref{x}. Note that when $x$
approaches $1/b$, \emph{i.e.}, when $\theta$ approaches $\theta_{0}$,
$N_{\mathrm{tot}}$ diverges, as we would expect. In our approximation
that $x\left(\mathrm{Planck\; surface}\right)\simeq x\left(\mathrm{enter\; slow\; roll}\right)$,
$x\left(\mathrm{Planck\; surface}\right)=1/b$ is identified as
the roll-up trajectory discussed in Sec.~\ref{hilltop}.

\subsection{How many $e$-folds should we expect in cosine inflation?\label{HowManyNatural}}

From Eq.~\eqref{Nanalytic2}, we know, given a trajectory that intersects
the Planck surface with angular coordinate $\theta$, how many $e$-folds
that trajectory will ultimately undergo. As in Sec.~\ref{HowManyQuadratic},
we now turn to the question of how many $e$-folds we should expect
under the canonical measure \eqref{trajmeasure} on the space of trajectories.
As shown in Sec.~\ref{trajectoryspace}, we must first find the Hamiltonian-conserved
measure --- a measure whose density satisfies the condition \eqref{Euler}
--- on effective phase space.

In our $z$ coordinates \eqref{coords}, the $H=M_{\mathrm{Pl}}$
surface corresponds to 
\begin{equation}
z=\frac{M_{\mathrm{Pl}}f}{\Lambda^{2}}\gg1.
\end{equation}
As we have previously noted, had we used a different ultraviolet cutoff
$\Lambda_{\mathrm{UV}}$, other than $M_{\mathrm{Pl}}$, all of the
results that follow would be the same, with $M_{\mathrm{Pl}}$ replaced
by $\Lambda_{\mathrm{UV}}$, so our conclusions would not qualitatively
change. Taking the large-$z$ limit of Eq.~\eqref{speed2}, we have
\begin{equation}
\mathbf{v}\simeq-3\frac{\Lambda^{2}}{f}z^{2}\sin^{2}\theta\mathbf{\hat{z}}-3\frac{\Lambda^{2}}{f}z^{2}\sin\theta\cos\theta\boldsymbol{\hat{\theta}},
\end{equation}
which is identical to Eq.~\eqref{highzspeed} up to a multiplicative
factor. That is, we have turned the large-$H$ behavior of cosine
inflation into the large-$H$ behavior of quadratic inflation, through
the judicious choice of coordinates \eqref{coords}. The effective
phase space measure density therefore takes the same general form
\eqref{measurezlarge} and on physical grounds we can restrict to
the $\gamma=3$ case for the reasons discussed in Sec.~\ref{HowManyQuadratic},
so that the measure density becomes as in Eq.~\eqref{sigmaPlanck}.

There is nevertheless an important difference between the quadratic
and cosine inflation scenarios, since $\phi$ is restricted to a small
window in the latter. This implies that the normalization of Eq.~\eqref{sigmaPlanck}
will be different. As we have noted, this restriction translates into
a restriction of values of $\theta$ on the Planck surface to a small
range near $\pi/2$ and near $3\pi/2$, with width given in Eq.~\eqref{x}.
Hence, $\sigma$ does not diverge on the $H=M_{\mathrm{Pl}}$ surface
for cosine inflation. In the $z$ coordinates, we have as in Eq.~\eqref{Hdot}
\begin{equation}
\dot{H}=-3\frac{\Lambda^{4}}{f^{2}}z^{2}\sin^{2}\theta.
\end{equation}
The probability distribution over the space of trajectories, parametrized
by the angle $\theta$ on the $H=M_{\mathrm{Pl}}$ surface, is therefore
\begin{equation}
\left.P\left(\theta\right)\right|_{H=M_{\mathrm{Pl}}}=\frac{3}{4\cos^{3}\theta_{0}}\left|\cos^{2}\theta\sin\theta\right|,\label{trajmeasure-2}
\end{equation}
with $\theta_{0}$ defined in Eq.~\eqref{x} as before. The normalization
once again comes from demanding that the total probability equal unity.

\begin{figure}
\noindent \centering{}\includegraphics[height=2in]{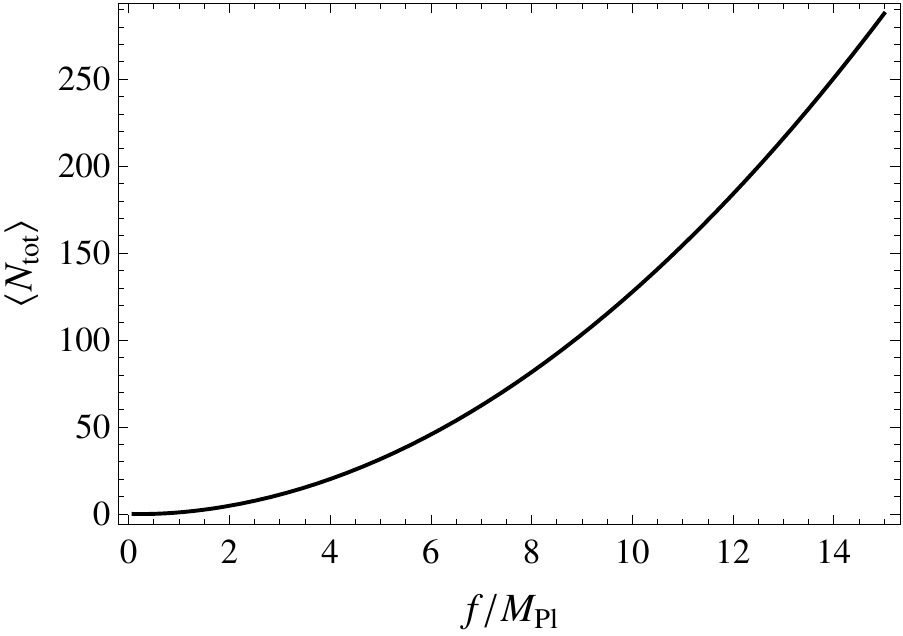}\caption{\label{Ntotnatural}Expected number of $e$-folds $\left\langle N_{\mathrm{tot}}\right\rangle $,
as computed in Eq.~\eqref{Ntotal2} using the canonical measure on
the space of trajectories, for cosine inflation with potential $V(\phi)=\Lambda^{4}[1-\cos(\phi/f)]$.}
\end{figure}

Having found the canonical measure \eqref{trajmeasure-2} on the space
of trajectories, we can now use our $e$-fold counting \eqref{Nanalytic2}
to compute the expectation value for the total number of $e$-folds
attained by a FRW universe in the cosine inflation model: 
\begin{equation}
\begin{aligned}\left\langle N_{\mathrm{tot}}\right\rangle  & =2\int_{\theta_{0}}^{\pi-\theta_{0}}N_{\mathrm{tot}}\left(\theta\right)\left.P\left(\theta\right)\right|_{H=M_{\mathrm{Pl}}}\mathrm{d}\theta\\
 & =\frac{9}{2b^{2}}\int_{0}^{1}\ln\left[\frac{1}{\left(1-u^{2}\right)\left(1+\frac{1}{3}b^{2}\right)}\right]u^{2}\mathrm{d}u\\
 & =\frac{f^{2}}{3M_{\mathrm{Pl}}^{2}}\left[8-6\ln2-3\ln\left(1+\frac{M_{\mathrm{Pl}}^{2}}{2f^{2}}\right)\right]\\
 & \simeq\left(\frac{8}{3}-2\ln2\right)\frac{f^{2}}{M_{\mathrm{Pl}}^{2}},
\end{aligned}
\label{Ntotal2}
\end{equation}
which is plotted in Fig.~\ref{Ntotnatural}; the constant $b$ was
defined in Eq.~\eqref{b}.%
\footnote{An exact expression for $H{\mathrm{d}}t$ \eqref{Hdtnatural} would
have $\varepsilon$ in place of $\epsilon_{V}$. Taking into account
relaxation of the slow-roll conditions near the hilltop, one can show
that, if $f\ll M_{\mathrm{Pl}}$, then the total $e$-fold count we
estimate should be increased by a factor of at most $M_{\mathrm{Pl}}/\sqrt{6}f$.
However, this would still lead to less than one $e$-fold of inflation
expected under the canonical measure in the $f\lesssim M_{\mathrm{Pl}}$
case. Moreover, one can show that, even near the hilltop, the approximation
$\epsilon_{V}\simeq\varepsilon$ is very accurate in the $f\gtrsim M_{\mathrm{Pl}}$
case.\label{approx}%
} For example, setting $f=1.2\times10^{19}$\,GeV (the unreduced Planck
mass) gives 
\begin{equation}
\left\langle N_{\mathrm{tot}}\right\rangle =32.
\end{equation}
This implies an insufficient amount of inflation to address the horizon
problem, but clearly $\left\langle N_{\mathrm{tot}}\right\rangle $
can be increased by a small boost in $f$.

Interestingly, $\left\langle N_{\mathrm{tot}}\right\rangle $ is independent
of $\Lambda$, only depending on $f$. This can be understood as follows:
for small $\phi\ll f$, cosine inflation is equivalent to quadratic
inflation, with mass $\Lambda^{2}/f$ taking the place of $m$. Then
the expected number of $e$-folds should be of order $N_{\mathrm{max}}$
\eqref{Nmax}, multiplied by a factor of $\cos^{2}\theta_{0}$, as
given in Eq.~\eqref{x}, to account for the limited allowed range
of $\phi$, \emph{cf.} Eq.~\eqref{Nanalytic}; this reasoning would
lead one to expect $\left\langle N_{\mathrm{tot}}\right\rangle \sim f^{2}/M_{\mathrm{Pl}}^{2}$,
which is indeed what we find. We find that $f>6.3\, M_{\mathrm{Pl}}=1.5\times10^{19}$
GeV is needed in order to have $\left\langle N_{\mathrm{total}}\right\rangle >50$.

Recently, in light of results from Refs.~\cite{Bicep2,Planck}, much
attention has been devoted to cosine inflation. By varying $f/M_{\mathrm{Pl}}$,
a one-parameter family of predictions is obtained that is able to
achieve agreement with either the Planck or BICEP2 results \cite{Freese}.
In particular, $f\sim5-10\times M_{\mathrm{Pl}}$ was found to be
in better agreement%
\footnote{Note that this range of $f$ could also be written as $f\sim1-2\times m_{\mathrm{Pl}}$,
where $m_{\mathrm{Pl}}=1/\sqrt{G}=1.2\times10^{19}$\,GeV is the
unreduced Planck mass.%
} with the Planck observations \cite{Planck}, while larger $f$ (which
brings the predictions closer to those of quadratic inflation) is
in better agreement with BICEP2.

What we have found is that smaller values of $f$ are, in the sense
of the canonical measure, highly unlikely to give a universe consistent
with the observed uniformity of the CMB. In particular, if $f\leq M_{\mathrm{Pl}}=2.4\times10^{18}$\,GeV,
we have less than one $e$-fold of inflation. More quantitatively,
we can compute the probability of attaining a given number $N_{0}$
of $e$-folds as a function of $f/M_{\mathrm{Pl}}$. From our expression
\eqref{Nanalytic2} for $N_{\mathrm{tot}}$ as a function of $\theta$
on the Planck surface, we find that this is just the probability that
\begin{equation}
\cos^{2}\!\theta\!>\!\cos^{2}\!\theta_{0}\!\!\left[1\!-\!\frac{3}{3+b^{2}}\exp\left(\!-\frac{2b^{2}}{3}N_{0}\!\right)\!\right]\!\equiv\cos^{2}\!\delta.
\end{equation}
That is, evaluating the integral $\mathrm{Pr}(N_{\mathrm{tot}}>N_{0})=4\times(3\sec^{3}\theta_{0}/4)\int_{\theta_{0}}^{\delta}\cos^{2}\theta\sin\theta\mathrm{d}\theta$,
we find 
\begin{equation}
\begin{aligned} & \mathrm{Pr}\left(N_{\mathrm{tot}}>N_{0}\right)\\
 & \;=\!1-\left[1\!-\!\left(1+\frac{M_{\mathrm{Pl}}^{2}}{2f^{2}}\right)^{-1}\!\exp\left(-\frac{M_{\mathrm{Pl}}^{2}}{f^{2}}N_{0}\right)\right]^{3/2}.
\end{aligned}
\label{probnatural}
\end{equation}

\begin{figure}
\noindent \begin{centering}
\includegraphics[height=2in]{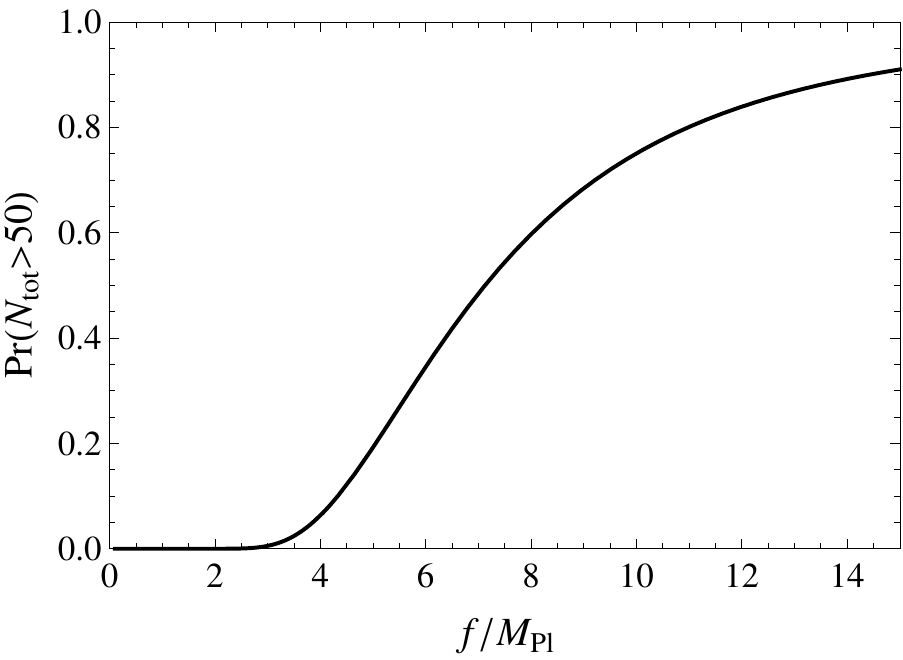} 
\par\end{centering}

\caption{\label{Ntotnaturalprob}The probability of obtaining 50 or more $e$-folds
of inflation as a function of $f$ for cosine inflation with potential
$V(\phi)=\Lambda^{4}[1-\cos(\phi/f)]$, as computed using the canonical
measure on the space of trajectories and starting on the $H=M_{\mathrm{Pl}}$
surface.}
\end{figure}

The result is plotted in Fig.~\ref{Ntotnaturalprob} for $N_{0}=50$.
We find that, if $f\leq2M_{\mathrm{Pl}}=4.9\times10^{18}$\,GeV,
the probability under the canonical measure of attaining 50 or more
$e$-folds of inflation is less than $10^{-5}$. While the details
of Eq.~\eqref{probnatural} break down if $f\lesssim M_{\mathrm{Pl}}$
due to corrections to the slow-roll approximation near the hilltop,
the expected total number of $e$-folds \eqref{Ntotal2} remains valid
and the probability of attaining more than 50 $e$-folds for $f\lesssim M_{\mathrm{Pl}}$
remains infinitesimal. That is, for $f\lesssim M_{\mathrm{Pl}}$,
the overwhelming majority of universes will under-inflate. On the
other hand, if $f=100M_{\mathrm{Pl}}=2.4\times10^{20}$\,GeV, we
find that the probability of a universe attaining at least 50 $e$-folds
of inflation is approximately $0.99964$. Hence, the probability of
a FRW universe undergoing sufficient inflation to explain the observed
uniformity of the CMB is sensitively dependent on $f/M_{\mathrm{Pl}}$
in cosine inflation, with larger values of $f\gtrsim\mathcal{O}\left(\mathrm{few}\right)\times M_{\mathrm{Pl}}\sim10^{19}$\,GeV
much preferred.

If $f$ is too small in cosine inflation, then our Universe is finely
tuned from the perspective of the canonical measure. Hence, models
of cosine inflation with $f$ on the order of $10^{18}$\,GeV or
less do not solve the cosmological fine-tuning problems that are the
original purpose of inflationary theory. For cosine inflation to truly
be natural in the cosmological sense, $f$ must be above $10^{19}$\,GeV.
On the other hand, this result helps motivate the possibility that
our Universe did experience just the right amount of inflation but
not too much, suggesting that there may be observable relics of the
pre-inflationary Universe that might be observable on very large angular
scales. Interestingly, in the large-$f$ limit that is favored by
the canonical measure, the observational predictions of cosine inflation
merge with those of quadratic inflation.

As with the quadratic case, we expect our results for cosine inflation
to be indicative of a more general lesson for hilltop (small-field)
models. Unlike the large-field case, where the potential rises all
the way to the Planck density, in cosine inflation the maximum is
well below that scale. There are trajectories that linger for an arbitrarily
large number of $e$-folds in the slow-roll regime near the top of
the hill, but there are also trajectories that exhibit a kinetic-dominated
phase of evolution prior to a finite period of slow roll. Our result
shows that it is the latter category that are most likely, as quantified
by the conserved measure on effective phase space.

\section{Conclusions}

The recent BICEP2 discovery, if verified, suggests that high-scale
cosmic inflation is the correct theory of the very early Universe.
With characteristic energy of order $10^{16}$\,GeV, observational
signatures of inflation open the door to physics on the threshold
of the Planck scale. Many models of inflation are currently being
investigated for their ability to fit precision CMB observations.
The current success of relatively simple models of inflation, driven
by a single scalar field with a potential and a canonical kinetic
term, is impressive. Given the large set of possible inflaton potentials,
it is of vital importance to develop useful theoretical tools that
enable observations to discriminate among competing models.

The theory of cosmic inflation was originally posited to solve problems
of fine-tuning of initial conditions, such as the uniformity of the
CMB temperature, lack of observed monopoles, and smallness of curvature.
Given the current wealth of precise cosmological measurements, it
is well-motivated to apply the same question of genericness to various
proposed models of inflation. That is, given a particular model, does
it generically produce the observed properties of our Universe? In
particular, does it typically produce the requisite number of $e$-folds
($40-60$) to account for the uniformity of the CMB? Inherent in such
questions is the idea of a measure: a probability distribution on
the set of all possible FRW universes. Following GHS \cite{GHS},
in Ref.~\cite{attractors} we developed a formalism for constructing
such a measure on the subset of flat universes (on which the GHS measure
diverges).

In the present work, we investigated the behavior of the effective
phase space measure for two general classes of potentials important
for single-field inflation: slow roll down a potential and lingering
behavior near a potential hilltop. In the former case, we showed that
the effective phase space measure generically becomes large, while
in the latter case it generically becomes small. That is, trajectories
that linger arbitrarily long near quasi-de Sitter space at a potential
hilltop are disfavored by the canonical measure, while trajectories
that slow roll and eventually reheat are favored. We next quantitatively
examined the statistical conclusions offered by the canonical measure
for two representative inflaton potentials: quadratic inflation and
cosine inflation. Interestingly, the statistical expectation for the
amount of inflation experienced in these two cases differed dramatically.
For quadratic inflation, we found that, given an inflaton mass consistent
with the observed amplitude of scalar perturbations \cite{Planck},
nearly all trajectories undergo 50 $e$-folds of inflation. In fact,
generic trajectories experience billions of $e$-folds. On the other
hand, for cosine inflation with symmetry-breaking parameter $f$,
typical trajectories under-inflate unless $f\gtrsim10^{19}$ GeV.
Above this scale, 50 $e$-folds are generically attainable and the
observational cosmology predictions of cosine inflation merge with
those of the quadratic potential.

From our demonstration with these two examples, we illustrated the
utility of the canonical measure in elucidating differences in physical
predictions among models of inflation. While a given potential may
have \emph{some} trajectory --- some possible history of a FRW universe
--- that undergoes enough inflation to correspond to our observed
Universe, that does not mean that this trajectory is generic. Indeed,
in some models, such as cosine potentials with $f\lesssim M_{\mathrm{Pl}}$,
the vast majority of trajectories, as weighted by the canonical measure,
do not undergo sufficient inflation, despite the existence of a small
subset of finely-tuned trajectories that do. The canonical measure
allows one to quantify the amount of tuning required in a given model
to reproduce our Universe.%
\footnote{All of these statements are made under the assumption of homogeneity,
ignoring perturbations. Given a universe in which inflation occurs
at all, this is a very good approximation.%
} The degree of tuning required on the space of trajectories to produce
at least 50 $e$-folds of inflation (or whatever other observed quantity
one is computing) should correspond inversely with the degree of credence
given a particular model, modulo theoretical bias. That is, given
two potentials, one generically attaining many $e$-folds and another
in which only a small subset (as computed in the canonical measure)
of trajectories attain 50 $e$-folds, the former model should be favored:
one could say that such a model is more ``natural,'' in the sense
that it requires less fine-tuning to match observations. This approach
is an interesting parallel to current discussions in particle physics
regarding naturalness of the electroweak scale and the amount of tuning
required in various models, such as supersymmetry.

The contrapositive of this line of thinking is also illuminating.
If future cosmological observations point to a particular inflaton
potential for which our Universe is not generic under the canonical
measure, that would shed light on even higher-scale physics. Such
a circumstance would tell us that our Universe is tuned --- on a non-generic
trajectory --- from the point of view of the classical measure. This
would indicate the importance of intrinsically quantum gravitational
processes or some ultimate theory of initial conditions.

As we enter an era of precision inflationary cosmology, models of
inflation will be subjected to increasingly refined measurement. In
the effort to determine which models best reflect reality, the notion
of naturalness, in the sense of genericness under the canonical measure
on the space of trajectories, can be very useful. The methods developed
in this work provide for quantitative probabilistic comparison among
models of inflation, providing a new means of shedding light on the
earliest moments of our Universe. 
\begin{acknowledgments}
This material is based upon work supported by the National Science
Foundation Graduate Research Fellowship Program under Grant No. DGE-1144469,
by DOE Grant No. DE-SC0011632, and by the Gordon and Betty Moore Foundation
through Grant No. 776 to the Caltech Moore Center for Theoretical
Cosmology and Physics. G.N.R. is supported by a Hertz Graduate Fellowship
and a NSF Graduate Research Fellowship. 
\end{acknowledgments}
 \bibliographystyle{apsrevGNR}
\bibliography{efoldsbib}

\end{document}